%% file: paper.tex
\documentclass[letterpaper,twocolumn,10pt]{styles/z2-article}
\usepackage{styles/usenix2019_v2,epsfig,endnotes}
\usepackage{hyperref}
\usepackage{mathptmx} 
\newcommand{\ignore}[1]{}
\usepackage{fancyhdr}
\usepackage[normalem]{ulem}
\usepackage{pgfplots}
\usepackage{comment}

\usepackage{amsmath}
\usepackage{upgreek}
\usepackage{graphicx}
\usepackage{color}
\usepackage{comment}
\usepackage{caption}
\usepackage{array}
\usepackage{booktabs}
\usepackage{algorithm}
\usepackage[noend]{algpseudocode}
\usepackage{amssymb}
\usepackage{cite}
\usepackage{paralist}
\usepackage[normalem]{ulem}
\usepackage[labelfont=bf]{caption}
\usepackage{subcaption}
\usepackage{textcomp}
\usepackage{titlesec}
\titlespacing*{\section}{0pt}{1\baselineskip}{\baselineskip}

\usepackage[export]{adjustbox}
\titlespacing*{\section}{0pt}{1\baselineskip}{\baselineskip}
\usepackage[export]{adjustbox}
\titlespacing*{\section}{0pt}{1\baselineskip}{\baselineskip}
\titlespacing{\section}{1pt}{*1}{*1}
\titlespacing{\subsection}{1pt}{*1}{*1}
\titlespacing{\subsubsection}{0pt}{*0}{*0}

\makeatletter
\renewcommand*{\p@section}{\S\,}
\renewcommand*{\p@subsection}{\S\,}
\def\tightitemize{\ifnum \@itemdepth >3 \@toodeep\else \advance\@itemdepth \@ne
\edef\@itemitem{labelitem\romannumeral\the\@itemdepth}%
\list{\csname\@itemitem\endcsname}{\setlength{\topsep}{-\parskip}\setlength{\parsep}{0in}\setlength{\itemsep}{0in}\setlength{\parskip}{0in}\def\makelabel##1{\hss\llap{##1}}}\fi}

\def\tightenumerate{%
\ifnum \@enumdepth >\thr@@\@toodeep\else
\advance\@enumdepth\@ne
\edef\@enumctr{enum\romannumeral\the\@enumdepth}%
\expandafter
\list
\csname label\@enumctr\endcsname
{\setlength{\topsep}{-\parskip}\setlength{\parsep}{0in}\setlength{\itemsep}{0in}\setlength{\parskip}{0in}\usecounter\@enumctr\def\makelabel##1{\hss\llap{##1}}}%
\fi}

\makeatother

\definecolor{red3}{rgb}{0.80,0.00,0.00}

\newcommand{\SK}[1]{\textcolor{red}{{#1}}}

\newcommand{\SKCITE}[1]{}

\usepackage{xspace}
\newcommand{\fsck}{\mbox{{e2fsck}}\xspace}
\newcommand{\Fsck}{\mbox{{E2fsck}}\xspace}

\newcommand{\xfs}{\mbox{\textsc{XFS}}\xspace}
\newcommand{\xfsck}{\mbox{xfs\_repair}\xspace}
\newcommand{\sysname}{\mbox{pFSCK}\xspace}
\newcommand{\pfscks}{\mbox{pFSCK-sched}\xspace}
\newcommand{\pfsckr}{\mbox{pFSCK-rsched}\xspace}
\newcommand{\ext}{\mbox{Ext4}\xspace}

\newcommand{\pipelinespliteq}{\mbox{pipeline-split-equal}\xspace}
\newcommand{\pipelinesplitbest}{\mbox{pipeline-split-optimal}\xspace}

\newcommand{\pfsckdatapara}{\mbox{pFSCK[datapara]}\xspace}
\newcommand{\pfsckpipeline}{\mbox{pFSCK[datapara+pipeline]}\xspace}
\newcommand{\pfsckpipelinespliteq}{\mbox{pFSCK[datapara+pipeline-split-equal]}\xspace}
\newcommand{\pfsckpipelinesplitbest}{\mbox{pFSCK[datapara+pipeline-split-optimal]}\xspace}

\newcommand{\CR}{C/R\xspace}
\newcommand{\CRs}{C/Rs\xspace}

\newcommand{\pfsckdataparallelfilepassoneperf}{2.1x\xspace} 
\newcommand{\pfsckdataparallelfilepasstwoperf}{1.8x\xspace} 
\newcommand{\pfsckdataparallelfileperf}{1.9x\xspace} 
\newcommand{\pfsckdataparalleldirpassoneperf}{1.8x\xspace} 
\newcommand{\pfsckdataparalleldirpasstwoperf}{1.3x\xspace} 
\newcommand{\pfsckdataparalleldirperf}{1.4x\xspace} 

\newcommand{\pfsckdatapipelinefileperfvsdata}{1.3x\xspace} 
\newcommand{\pfsckdatapipelinedirperfvsdata}{1.1x\xspace} 

\newcommand{\pfsckschedfileperfvspipe}{1.1x\xspace} 
\newcommand{\pfsckschedfileperf}{2.6x\xspace} 
\newcommand{\pfsckschedirperfvspipe}{1.05x\xspace} 
\newcommand{\pfsckscheddirperf}{1.6x\xspace} 
\newcommand{\pfsckschedperf}{2.6x\xspace} 

\newcommand{\pfsckschedperfSSD}{2.1x\xspace} 

\newcommand{\fsckfsckdeg}{1.2x\xspace} 
\newcommand{\fsckrocksdeg}{1.5x\xspace}  
\newcommand{\pfsckpfsckdeg}{1.07x\xspace} 
\newcommand{\pfsckrocksdeg}{1.05x\xspace} 

\newcommand{\fsckfsckdegonline}{1.4x\xspace} 
\newcommand{\fsckrocksdegonline}{1.6x\xspace}  
\newcommand{\pfsckpfsckdegonline}{1.2x\xspace} 

\newcommand{\pfskvsfsckcpusharingspeeduponline}{1.7x\xspace}

\newcommand{\totalpfsckoverxfs}{1.8x\xspace}
\newcommand{\totalpfsckoverfsck}{\pfsckschedperf}

\newcommand{\totalpfsckmemoverhead}{1.17x\xspace}

\newcommand{\nvmesysconfig}{64-core Dual Intel® Xeon Gold 5218, 2.30GHz, 64GB of DDR
memory, and 1TB NVMe Flash Storage running Ubuntu 18.04.1}

\begin{document}

\date{}

\title{Accelerating Filesystem Checking and Repair with pFSCK \vspace{-0.0in}}

\author{
{\rm David Domingo, Kyle Stratton, Sudarsun Kannan}\\
Rutgers University
}

\maketitle


\input{sections/0_abstract}
\input{sections/1_introduction}
\input{sections/2_motivation}
\input{sections/3_case}
\input{sections/4_design}
\input{sections/5_implementation}
\input{sections/6_evaluation}
\input{sections/8_conclusion}

{\footnotesize
\bibliographystyle{plain}
\bibliography{paper,paper2}
}

\end{document}

%% file: sections/0_abstract.tex
\vspace{-0.4in}
\begin{abstract}
File system checking and recovery (\CR) tools play a pivotal role in
increasing the reliability of storage software, identifying and correcting file
system inconsistencies. However, with increasing disk capacity and data
content, file system \CR tools notoriously suffer from long runtimes. We
posit that current file system checkers fail to exploit CPU parallelism
and high throughput offered by modern storage devices.

To overcome these challenges, we propose \sysname, a tool that redesigns C/R to
enable fine-grained parallelism at the granularity of inodes without impacting
the correctness of \CR's functionality. To accelerate \CR, \sysname first
employs data parallelism by identifying functional operations in each stage of
the checker and isolating dependent operation and their shared data structures.
However, fully isolating shared structures is
infeasible, consequently requiring serialization that limits scalability. To reduce the
impact of synchronization bottlenecks and exploit CPU parallelism, \sysname
designs pipeline parallelism allowing multiple stages of C/R to run
simultaneously without impacting correctness. To realize efficient pipeline
parallelism for different file system data configurations, \sysname provides
techniques for ordering updates to global data structures, efficient
per-thread I/O cache management, and dynamic thread placement across different
passes of a \CR. Finally, \sysname designs a resource-aware scheduler aimed
towards reducing the impact of \CR on other applications sharing CPUs and the file
system. Evaluation of \sysname shows more than \totalpfsckoverfsck gains of \fsck and
more than \totalpfsckoverxfs over XFS's checker that provides coarse-grained parallelism.

\end{abstract}

%% file: sections/1_introduction.tex
\section{Introduction}
Modern ultra-fast storage devices such as SSDs, NVMe, and byte-addressable NVM
storage technologies offer higher bandwidth capabilities and lower latency compared to
hard-disks providing better opportunities for exploiting CPU parallelism. While
I/O access performance has increased, storage hardware errors have continued to
grow coupled with newer and exploratory high-performance designs impacting file
system reliability~\SK{\cite{Gunawi:FailSlow, Alagappan:Correlated, btrfsbugs}}.
For decades, file system checking and repair tools (referred to as \CR
henceforth) has played a pivotal role in increasing reliability of software
storage stacks, identifying and correcting file system inconsistencies~\cite{mckusick1986fsck}. In
fact, in the event of a system crash or storage failure in data centers, file
system checkers are typically used as the first remedial solution to system
recovery~\cite{Gunawi:FailSlow}.

\par File system \CR tools work by identifying and fixing the structural
inconsistencies of file system metadata, such as inconsistencies in inodes, data
and inode bitmaps, links, and directory entries. Well-known and widely used
tools such as \fsck (file system checker for \ext) divides \CR across multiple stages (commonly referred to as
passes) with each pass responsible for checking a file system structure (e.g.,
directories, files, links).  However, \CRs are known to be
notoriously slow, showing a linear increase in \CR time with an increase in file
and directory count and the disk utilization~\cite{Henson:ext2fsck, mckusick1986fsck,
Lu:IncrementalChecker, McKusick:fsckFreeBSD, Ma:Ffsck}. Although modern flash
and NVM technologies provide lower latency and bandwidth, current \CR tools fail
to utilize such hardware capabilities or multicore CPU parallelism fully. While
modern \CRs have attempted to increase parallelism, their coarse-grained
approaches, such as parallelizing C/R across logical volumes or
logical groups, are insufficient to accelerate \CR on file systems with data
imbalance across logical groups~\cite{Gunawi:SQCK, Ma:Ffsck, Henson:ext2fsck,
Mtanski:XFS}.

To overcome such limitations, we propose \textit{\bf \sysname}, a parallel \CR
that exploits CPU parallelism and modern storage's high bandwidth to accelerate
file system checking and repair time without compromising correctness.
Accelerating file system C/R could significantly reduce system downtime and
improve storage availability~\cite{Gunawi:SQCK, Ma:Ffsck, Gunawi:FailSlow, Alagappan:Correlated}.  In this pursuit, \sysname
introduces fine-grained parallelism, i.e., parallelism at the granularity of
inodes and directory blocks, resulting in a significantly faster execution compared to
traditional \CRs. \sysname first employs {\bf data parallelism} by breaking up the
work done at each pass, redesigning data structures for scalability, and
allowing multiple threads to process.  Although data parallelism accelerates
checking, updates to global data structures (e.g., bitmap) within each pass
are designed to match the file system's layout (e.g., block bitmap in an \ext file
system) and must be synchronized and serialized to ensure checking correctness. As a
result, with increasing threads, the cost of synchronization and serialization
can quickly outweigh the performance gains.  Hence, \sysname introduces {\bf
pipeline parallelism} to parallelize \CR along with the logical flow (i.e.,
across multiple passes).

\par Supporting data and pipeline parallelism within \sysname requires addressing
several challenges. First, updates to data structures shared must be ordered for
\CR correctness. For example, a directory cannot be certified to be error-free
by the directory checking pass unless all its files are verified as consistent by the
inode checking pass. To address these ordering constraints, taking inspiration
from out-of-order executions in hardware processors, we isolate the global data
structures and perform all necessary operations in parallel but certify
correctness only when the results are merged. Second, static partitioning of CPU
threads across different passes is suboptimal because each pass checks
different metadata (e.g., file, directory, links) and the amount of each kind of
metadata can vary across different file system configurations. In addition, the time to
process different types of metadata vary significantly (e.g., checking a directory
can take significantly longer than a file). Hence, we propose a dynamic thread
scheduler that monitors progress across different passes of \sysname
and uses the pending work ratio for thread assignment.

\par Third, I/O optimizations such as I/O caching and read-ahead mechanisms in
current \CRs are not designed for multithreaded parallelism, which we address by
designing a thread-aware I/O caching, thereby substantially reducing I/O
wait-times. Finally, to exploit multi-core parallelism in ways that do not
affect the performance of other co-running applications that share CPUs or
access the same disks checked by \CR (online checking), we propose a
resource-aware mechanism that allows for scaling the number of threads used to
perform checking by monitoring the overall CPU utilization of the system.

\par The combination of \sysname's above techniques significantly reduces \CR
runtime. For example, \sysname's data parallelism and pipeline parallelism on an
800GB file system on NVMe reduce runtime by up to \pfsckschedfileperf for a
file-intensive disk configuration. For a directory-intensive disk configuration,
\sysname is able to reduce runtime by up to \pfsckscheddirperf compared to \fsck
and by up to \totalpfsckoverxfs over the XFS file system checker.  In the
pursuit of increasing multicore parallelism, \sysname increases memory usage by only
\totalpfsckmemoverhead over \fsck (from 3~GB in \fsck to 3.5~GB in \sysname
for a 800GB file system).  Further, \sysname's scheduler increases gains by
\pfsckschedfileperfvspipe over \sysname without a scheduler. When sharing the
CPUs between \sysname and RocksDB, the system-wide resource-aware mechanism
minimizes \sysname performance degradation to \pfsckpfsckdeg as well as limits
the overhead on RocksDB by \pfsckrocksdeg. Finally, \sysname provides a
significant performance boost during live checking compared to \fsck improving
performance by \pfskvsfsckcpusharingspeeduponline.

%% file: sections/2_motivation.tex
\section{Background and Motivation}
\label{sec:motivate}
Storage hardware advancements have opened up the potential for accelerating I/O
bound applications. One critical set of applications that could potentially
benefit are the file system checking and repair (\CR) tools, which run in almost
all computing systems. We first give some background on current hardware trends,
\CR tools, and then discuss prior approaches that accelerate \CR and their
limitations.

\subsection{Hardware and Software Trends}
With increasing core count and the advent of faster storage devices, system
performance has seen vast speedups due to hardware advancements. More
specifically, flash memory technologies like PCI-attached SSD and NVMe devices
provide increased throughput (8-16 GB/s) and lower access latency (20-50
$\mu$s) compared to hard disks. At the other
end, fast storage class memories such as Intel’s DC Persistent
Memory~\cite{intel:3dxpoint} and other classes of nonvolatile memory (NVM)
directly attach to the memory controller and provide byte-addressable
persistence. These technologies scale 4x larger than DRAM capacity, with
variable read (100-200ns) and write (400-800ns) latencies, and bandwidth
capabilities ranging from 8 GB/s to 20 GB/s.

On the software side, a huge body of prior research is in progress to redesign
and optimize file systems for modern storage hardware, which includes file
systems for SSD, NVMe, and NVMs ~\cite{F2FS, ou2016high, kadekodi2019splitfs,
NOVA, Kwon:Strata} and storage stack in general~\cite{Kannan:NoveLSM:ATC18, DBLP:conf/hipc/FernandoKGS16}.
Besides, the open-source community is investing a
substantial effort to optimize traditional file systems such as \ext and \xfs for
modern storage hardware, given the wide-usage and reliability of these file
systems. For example, file systems such as \ext-DAX continue to retain
traditional file system structure and optimize performance by removing
components such as page cache, schedulers, and logging. Reducing data corruption
for both these approaches can be challenging and requires a few years of
production use ~\cite{jaffer2019evaluating, aghayev2019file}.

\subsection{File System Checking and Repair}
Since the dawn of file systems, file system consistency has always been an
issue\SKCITE{~\cite{ArpaciDusseau14-Book, Pillai:CCFS, Pillai:Alice}}. Though
modern file systems deploy mechanisms such as journaling, copy-on-write,
log-structured writes, and soft updates to handle inconsistencies, they cannot fix
errors that may have been present due to corruptions manifested in the past by events
such as a failing disk, bit flips, overheating, or correlated
crashes~\cite{Jaffer:Reliability,bairavasundaram2008analysis,bairavasundaram:latenterrors,chidambaram2013optimistic,
Zheng:2016:RAS:3014162.2992782}. A widely-used approach to handle disk and file
system corruptions and errors is to check and fix inconsistencies using file
system \CR tools, such as \fsck and \xfsck, that scan file system metadata for
corruptions and fix them. Most \CR tools are designed specifically for the
layout of a file system. \CRs such as \fsck and \xfsck have multiple passes that
check inode consistency, directory consistency,
connectivity of all directories, directory entries, and lastly,
the reference counts of inodes and blocks. The repair process involves fixing
errors such as updating the block bitmap that does not indicate a block referenced
by a file as being used. For more complex
errors (e.g., a block referenced by several inodes), administrators are given
an option to accept or decline repairs.

\subsection{Related Work}
With increasing disk capacities and file system size, \CRs notoriously tend to
run longer; specifically, the increase in system downtime may dominate the repair cost by
orders of magnitude
~\cite{proactivereplica, linuxqfsck, redhatfsck, stackexchangefsck, Henson:ChunkFS}.
We next discuss the state-of-the-art C/R optimizations for offline
(unmounted file systems) and online C/Rs and their limitations.

\par \noindent{\bf Offline \CRs. }
To reduce C/R time, open-source \CRs, such as the \ext's \fsck and XFS file
system's \xfsck, parallelize checking across disks (\fsck) or logical groups
(\xfsck). C/R techniques like Ffsck~\cite{mckusick1986fsck} and
Chunkfs~\cite{henson2006chunkfs}, speed up by modifying the file system to
provide a better balance across logical groups. For example, Chunkfs is designed
to utilize disk bandwidth by partitioning the file system into smaller, isolated
groups that can be repaired individually and in parallel, whereas Ffsck
~\cite{mckusick1986fsck} rearranges metadata blocks within the file system to
reduce seek cost and optimize file system traversal.  SQCK~\cite{gunawi2008sqck}
enhances \CR by utilizing declarative queries for consistency checking across
file system structures. Overall, while prior C/R designs have attempted to
improve file \CR performance, they suffer from several weaknesses. First,
Chunkfs (and XFS) require a coarse grain separation of file system blocks to
accelerate file system checking. Take the case of an imbalanced file system with
several large files spread across different logical groups of a disk.
Interestingly, we show in Section~\ref{sec:eval} the \xfsck's parallelism is
limited to simple inode scans, omitting any parallelism for checking directory
metadata. Other techniques such as SQSCK and Ffsck require intrusive changes to
the way we manipulate file system metadata or need completely rebuilding \CR,
which could reduce or prevent widespread adoption.

\par \noindent{\bf Online \CRs. }
To reduce system C/R downtime, proprietary online \CRs such as WAFL file
system's Iron~\cite{Ram:WaflIron} (a NetApp-based \CR tool for WAFL file system)
and ReFS~\cite{Gawatu:ReFS} fix corruptions as they are encountered allowing
file system operations to continue. WAFL-Iron performs incremental live \CR.
Because storage blocks are made available as \CR is in progress, WAFL-Iron
imposes invariants such as (1) checking all blocks before any software use, (2)
checking ancestor blocks (directory) before any data or metadata block (inode
block) is checked. These invariants avoid repeated checking of an inode for
every data block, and also reduce memory usage. To scale C/R to petabytes,
WAFL-Iron expects the presence of block-level checksums, RAID, and most importantly,
good storage practices by customers. Open-source C/Rs such as \fsck allows for
online checking by utilizing LVM-based snapshotting and running \CR on the
snapshot while the file system is still in use ~\cite{e2scrub}. We
evaluate \fsck's LVM-based online C/R in Section~\ref{sec:eval}. Recon protects
file system metadata from buggy operations by verifying metadata consistency at
runtime~\cite{fryer2012recon}. Doing so allows Recon to detect metadata
corruption before committing it to disk, preventing error propagation. Recon
does not perform a global scan and hence cannot identify or fix errors
originating from hardware failures.

\par \noindent{\bf \CR Correctness. }
To ensure the correctness and crash-consistency of C/Rs itself and recover more
reliably in light of system faults, Rfsck-lib~\cite{Gatla:Robust} provides C/Rs
with robust undo logging.  \sysname's fine-grained parallelism goals are
orthogonal to Rfsck-lib, however, incorporating Rfsck-lib can improve the reliability
of \sysname in case of failures.

\textbf{Summary. }
To summarize, unlike prior systems, \sysname is aimed towards fine-grained
parallelism, the ability to utilize storage bandwidth efficiently across
multiple passes of \CR, adapting to system resources, and the capability to
reduce the impact on other applications.

%% file: sections/3_case.tex
\section{Motivation and Analysis}
\label{sec:case}
In the pursuit of accelerating C/Rs, we first decipher the performance
bottlenecks of the widely-used \ext file system's \fsck C/R tool. We first
provide an overview of \fsck and then examine \fsck's runtime for
different file system configurations. For brevity, we study \xfsck in
Section~\ref{sec:eval}.

\subsection{\Fsck Overview}
\Fsck uses five sequential passes for C/R: the first pass (referred to as
Pass-1) checks the consistency of inode metadata; Pass-2
checks directory consistency; Pass-3 checks directory connectivity; Pass-4
checks reference counts; finally, Pass-5 checks data and metadata bitmap
consistency.

\subsection{Analysis Setup}

\begin{figure}[t]
\hspace{-0.2in}
 \begin{subfigure}[t]{0.22\textwidth}
    \includegraphics[width=4.5cm,keepaspectratio]{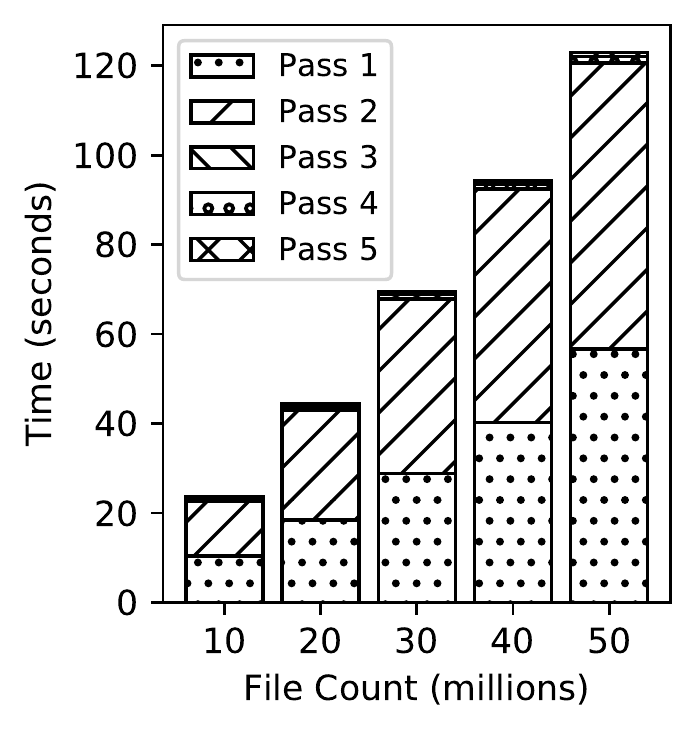}
    \captionsetup{justification=centering, margin={0.5cm,0cm}}
    \caption{\textbf{File Count Sensitivity:} \\ \footnotesize{Runtime of \fsck as \\ total file count increases}}
    \label{fig:filecount}
 \end{subfigure}
 \hspace{0.1in}
 \begin{subfigure}[t]{0.22\textwidth}
    \includegraphics[width=4.25cm,keepaspectratio]{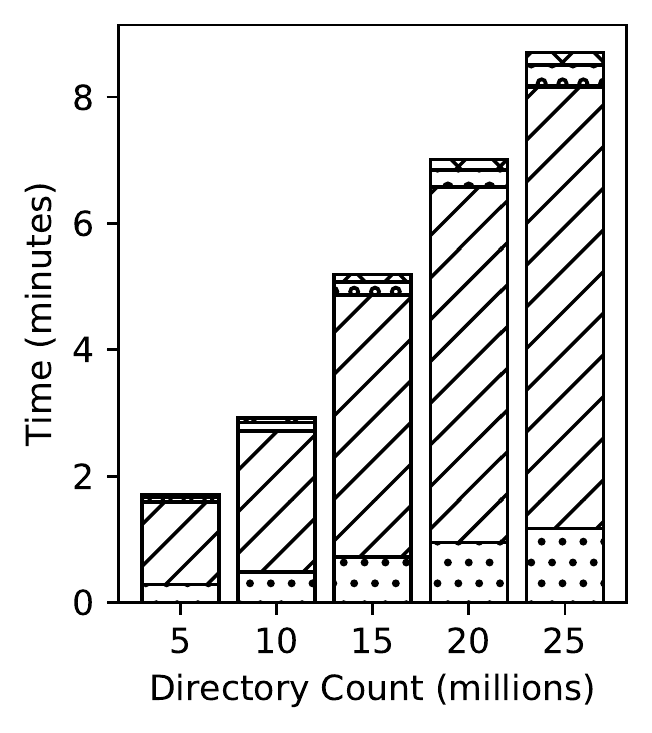}
    \vspace{-0.17in}
    \captionsetup{justification=centering}
    \caption{\textbf{Directory Count Sensitivity:} \footnotesize{Runtime of \fsck as total directory count increases}}
    \label{fig:dircount}
 \end{subfigure}
 \caption{\textbf{Runtime of C/R for an 800GB file system with varying counts of files or directories}}
\end{figure}

To analyze and decipher the breakdown of \fsck's runtime, we run \fsck on file
systems with varying configurations. We conduct our analysis on a \nvmesysconfig. We
fill the file system using \textit{fs\_mark}, an open-source, file system benchmark
tool~\cite{Wheeler:fsmark}. For our analysis, we mainly focus on file systems without
corruptions. 
To get a finer understanding of how \fsck scales with file system
configurations, we study the sensitivity of \CR's runtime for multiple file
system variables such as file count and directory count.

\begin{figure}[t]
\vspace{-0.0in}
 \includegraphics[width=8cm,keepaspectratio]{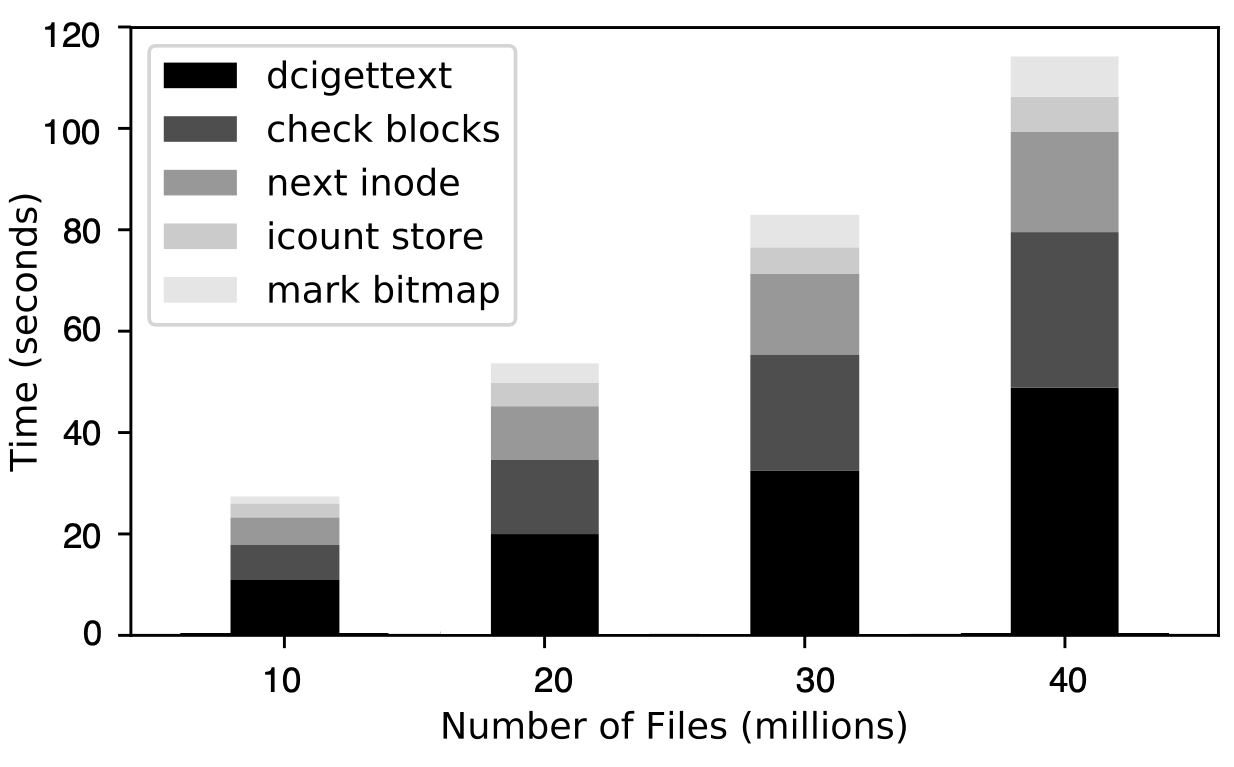}
 \vspace{-0.1in}
  \caption{\textbf{\fsck Pass-1 Time Breakdown\small{.}} \footnotesize{Time spent within inode checking
  pass (Pass-1) as the total file count increases.}}
 \vspace{-0.0in}
\label{fig:pass1breakdown}
\end{figure}

\noindent{\bf File-intensive file systems.}
First, to understand how file count affects runtime, we generate multiple file
intensive file system configurations with a 95:1 files to directories ratio.
Operating on file-intensive file systems, Pass-1, which checks the consistency
of inodes structures, dominates \fsck runtime, followed Pass-2, which checks
directory block consistency. Figure~\ref{fig:pass1breakdown} shows the
function-wise breakdown in Pass-1 that checks the
consistency of file inodes as well as track directory blocks encountered to be examined
in the next pass. We notice a function \texttt{dcigettext} (a
seemingly innocuous) language translator used for error handling gets
(incorrectly) used for every inode check and poses a substantial slowdown on the
C/R performance. \footnote{We reported this to \fsck developers, and the fix has
been upstreamed.} Other Pass-1 steps such as \texttt{check blocks} that checks the
blocks referenced by an inode, \texttt{next inode} that reads the next inode blocks from disk,
\texttt{mark bitmap} that updates global bitmaps to track the metadata encountered, and
\texttt{icount store} that stores inode references also increase in runtime.  Although the
number of directories is small, the Pass-2 (directory checking pass) runtime
increases because the number of directory blocks that store directory entries increase.
Pass-3 checks connectivity and ensures the reachability of directories from the
root. For a small directory count, the runtime is a small compared to the runtime of Pass-1 and
Pass-2. We also find that increasing file size while keeping the number of files
constant does not increase \fsck runtime significantly (not shown
for brevity).

\noindent{\bf Directory-intensive file system. }
We next analyze the impact of the directory count on the runtime of \fsck in
Figure~\ref{fig:dircount}. We generate multiple directory
intensive file systems, each with increasing number of directories. We define
a directory-intensive file system as a file system with over a 1:1 file to directory ratio.
To ensure that each directory requires the same amount of work, we create a single file in each directory.
As expected, with an increase in directory count, Pass-2's runtime significantly increases due to the increased
number of directory blocks as well as the fact that directory blocks hold checksums for consistency,
forcing Pass-2 to recompute and verify these directory blocks checksums. Interestingly,
Pass-1's runtime also increases; this is because Pass-1 is responsible for
identifying directory blocks and adding to a directory block list
(\textit{db\_list}) to be used within Pass-2. 

\begin{figure}[t]
\hspace{-0.2in}
 \begin{subfigure}[t]{0.22\textwidth}
 \includegraphics[width=4.5cm,keepaspectratio]{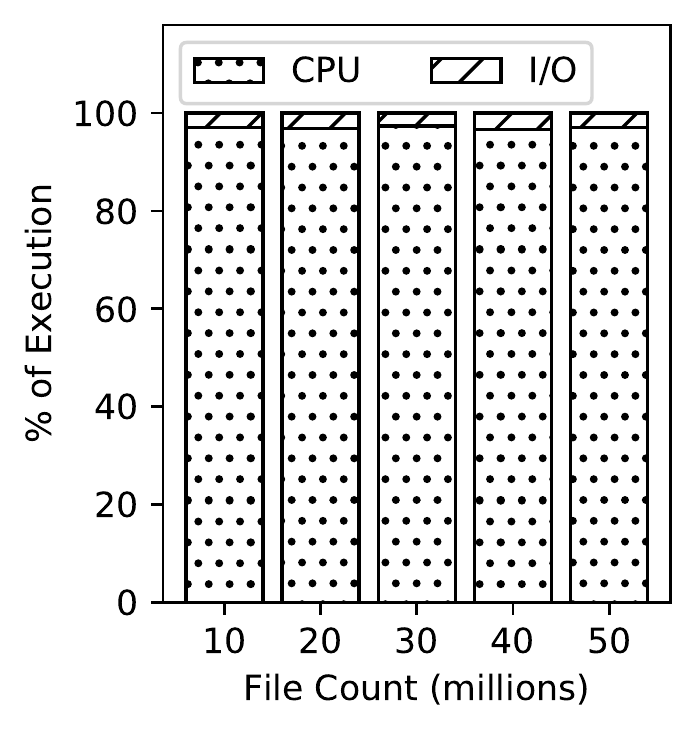}
 \label{fig:awarefiles}
 \end{subfigure}
\hspace{0.1in}
 \begin{subfigure}[t]{0.22\textwidth}
 \includegraphics[width=4.5cm,keepaspectratio]{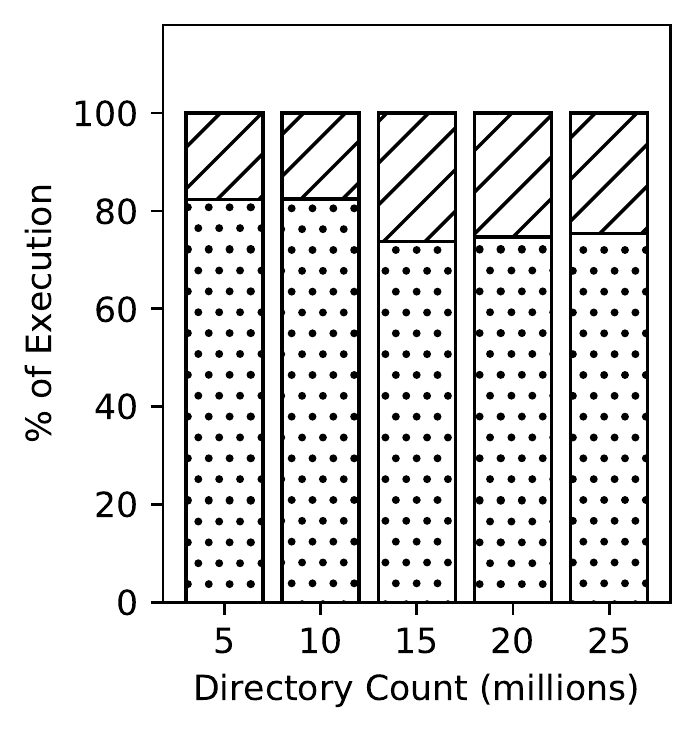}
 \label{fig:awaredirs}
 \end{subfigure}
 \vspace{-0.25in}
 \caption{\textbf{Percent of I/O wait time (Left: Files-intensive file system,
  Right: Directory-intensive file system)}}
 \label{fig:e2fsckio}
\end{figure}

\subsection{I/O utilization}
To understand the computational vs. I/O bottlenecks, in Figure
\ref{fig:e2fsckio}, we show the compute vs. I/O wait time ratio for \fsck. As
shown, the compute time dominates the I/O wait time. In all our experimental runs, we
observe that \fsck's peak and average I/O bandwidth usage is ~260~MB/s and
100~MB/s, respectively, on an NVMe device with 2~GB/s and 512~MB/s sequential
and random read bandwidth. \textit{In general, file system C/R tools (e.g., \fsck
and XFS\_repair) not only suffer from I/O access time but also computational
cost.}

\textbf{Summary. } To summarize, our analysis shows high runtime overheads of
\fsck across file system configurations; this is mainly due to the
serial, single-threaded nature of \fsck, designed in the era of spinning hard
drives. The linear complexity of its runtime is unsuitable as file system
capacities trends upward, potentially taking hours, or even days, to check datacenter-scale
file systems. Besides, C/R's repair when there are file system inconsistencies
could further increase \CR runtime.

%% file: sections/4_design.tex
\section{Goals and Design Insights}
\label{sec:design}
\sysname aims to overcome the limitations of traditional file system C/Rs by
exploiting fine-grained multi-core parallelism, higher disk bandwidth, efficient
use of CPUs using a \sysname scheduler, and ways to reduce the impact on other
co-running applications. We next outline the goals and provide \sysname's design overview.

\subsection{\sysname Goals}
\textbf{Decrease file system \CR runtime. }
The main goal is to make file system \CR faster. We want to increase the
speed at which file system metadata can be scanned and inconsistencies
identified, without compromising repairing capabilities.

\textbf{Adapt to different file system configurations.}
The \CR performance should improve regardless of file system size, utilization,
or configurations, such as a file-intensive or directory-intensive file system.

\textbf{Support offline and online C/R.}
\CRs can be used when a disk is not mounted, and the file system is offline or
when a system is online, and the file system is actively used. Hence, \sysname
aims to support both offline and online \CR.

\textbf{Adapt to system utilization.}
\CR should have the ability to adapt to varying system resource utilization over
time to reduce the potential performance impact on any currently-running
applications. \sysname aims to adapt to varying system-wide CPU use.

\subsection{\sysname Design Insights}
 We next describe the key design insights to realize the above goals.

\textbf{Insight 1: Maximize potential bandwidth through multiple cores and data
parallelism.}
To overcome the bottlenecks of current serial \CRs and \CRs that parallelize
at a coarse granularity such as across logical volumes or logical groups, \sysname exploits
fine-grained inode and directory block parallelism. Towards this goal,
\sysname first introduces data parallelism to \CR for better utilization of CPU
parallelism enabled by modern storage bandwidth capabilities. At a high-level,
in each pass, basic file system structures such as inodes, directory blocks, dirents, and
links are divided and checked across a pool of worker threads. While
seemingly simple, achieving data parallelism requires data structure isolation
across threads to reduce synchronization bottlenecks.

\textbf{Insight 2: Enable pipeline parallelism by reducing inter-pass dependencies.}
Though data parallelism improves performance, updates to several inter-pass
global data structures used for building a consistent view of the file system
and identifying inconsistencies (ex. bitmaps), must be serialized.
Consequently, this limits data parallelism's performance capability with higher
CPU counts and also degrades performance at higher thread counts due to
contention on the shared structures. Pipeline parallelism breaks the rigid wall across passes allowing
multiple passes to be executed simultaneously along with the logical flow of an
application, thereby increasing CPU parallelism and reducing the performance
impact of serialization. To realize pipeline parallelism requires managing
per-pass thread pools, isolating inter-pass shared structures using divide and
merge approaches, delineating checking and certification of inodes, and reducing
I/O wait times.

\textbf{Insight 3: Adapt to file system configurations with dynamic thread scheduling.}
Enabling data and pipeline parallelism requires assigning threads across
different passes of \sysname. Static partitioning of CPU threads across
different passes are suboptimal due to lack of information about metadata types
(files, directories, links) and work across passes; for example, checking directory
blocks in Pass-2 (directory checking pass) require more processing time than checking inodes in Pass-1 (file checking pass)
as discussed in Section~\ref{sec:case}. To overcome the challenge of
accelerating \CR for different file system configurations, we design a dynamic
thread scheduler that assigns threads to process different types of file
system objects as they are discovered and migrates threads across different
passes of the pipeline.

\textbf{Insight 4: Reduce system impact through resource utilization
awareness.}
File system C/Rs could potentially run with other applications sharing CPUs while
performing checking on separate disks. Given \sysname's goal to exploit
available CPUs, this could potentially impact other co-running applications.
Similarly, \CR could run on disks that are also actively used by other
applications to store data. To reduce the overall system impact on co-running
applications as well as \sysname, we equip \sysname's scheduler with resource
awareness to dynamically identify the number of cores to use at any single point
in time to minimize potential impact on other co-running applications without significantly
impacting \sysname's performance.

%% file: sections/5_implementation.tex
\section{Design and Implementation}
To realize the goals of \sysname, we discuss the design and implementation of
\sysname's data parallelism, pipeline parallelism, dynamic
thread scheduler, and resource-aware scheduling. \sysname extends \fsck to
realize these design changes.

\subsection{Data Parallelism}
\sysname's data parallelism divides work in each pass among a group of worker
threads on the granularity of inodes and enables concurrent C/R. While seemingly
simple, efficient data parallelism during C/R demands an efficient threading
model for fine-grained inode parallelism, functional separation of C/R within
each pass, and per-thread contexts for isolating data structures and reducing
synchronization cost.

\noindent{\bf Fine-grained Inode-level Parallelism.}
For fine-grained inode-level parallelism, \sysname uses the superblock
information to identify the total number of inodes in the file system and evenly
divides the inodes across a given set of C/R workers. To reduce the cost of
worker threads management, \sysname uses a thread-pool framework
\cite{Qin:Arachene} that provides the ability to assign tasks to
multiple worker threads. The worker threads are then reused across different
passes of a \CR. \sysname also co-locates threads of a pass to the same CPU and
memory socket to avoid the lock variable bouncing across processor caches on
different sockets. We will also discuss the need for dynamically identifying
work done across threads and scheduling in Section~\ref{dynamicsched}.

\noindent{\bf Functional Parallelism for Reducing Synchronization Overheads.}
Only dividing inodes for checking across worker threads is insufficient. To
benefit from fine-grained parallelism, it is critical to reduce synchronization
across worker threads in each pass without compromising correctness.

\par We first break each \CR pass into four main functional steps and reduce
synchronization across these steps. The steps include: (1) file system metadata
C/R, (2) global file system metadata update, (3) \CR-level accounting, and
finally, (4) intermediate result sharing; these four steps comprise 95\% of the
work. The metadata check performs logical checks that verify their integrity
across each pass (for example, blocks of an inode). Next, updating global file
system metadata includes updating file system-level bitmaps that keep track of
blocks and inodes currently used and referenced. The bitmaps are also used to
detect any inconsistencies between inodes such as duplicate block references
where more than one inode claim the same block. Third, \CR-level accounting
involves updating counters that track statistics such as file types. Finally,
intermediate result sharing across passes involves creating and updating data
structures such as a red-black tree with inode information and a hash-tree based
directory list.

\par While synchronization between file system metadata check (step 1)  and
global metadata update steps (step 2) are essential, synchronization between
first two steps and step 3 (\CR counter/statistics update) can be avoided by
allowing threads to maintain per-threads stats. The results of step 1 and 2 can
be aggregated before the next pass, reducing the synchronization cost
significantly.

\noindent{\bf Thread Contexts for Isolation.}
In current \CRs such as \fsck, upon which \sysname is built, we identify
significant data structure sharing across functions (steps 1 to 4) inside each
pass and across passes. To reduce sharing
and provide isolation, we introduce per-thread contexts (in contrast to a global
context in \fsck). The thread contexts are similar to OS thread contexts and
contain information such as buffers used for processing file system objects,
intermediate data structures, structures to track progress, locks held for shared
data structures, and CPUs used. At the end of each pass, the information
within each thread context, such as per-thread buffers and generated intermediate data structures, are
aggregated before the subsequent pass.

\begin{figure}
 \center{\includegraphics[width=8.5cm]{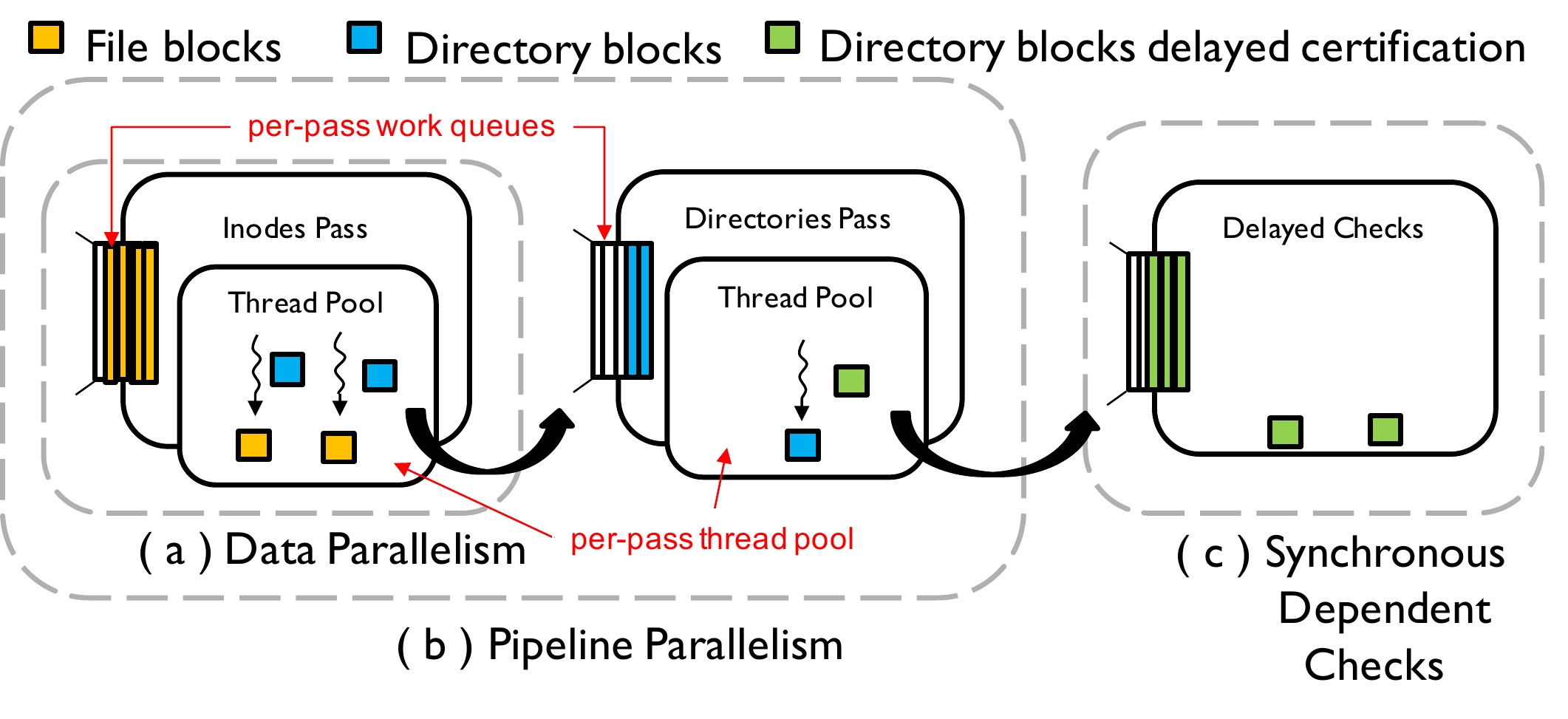}}
  \vspace{-0.3in}
  \caption{\textbf{Parallelism in \sysname.}
  \small{(a) Thread pools within each pass allows for data to be operate in parallel (\textit{data parallelism}).
  (b) Use of multiple thread pools allows each pass of \sysname to operate simultanously (\textit{pipeline parallelism}).
  (c) Any dependent checks needed to be carried out synchronously is delayed
  within its own logical pass.}}
 \label{datafull}
\end{figure}

\subsection{Pipeline Parallelism}
While data parallelism achieves concurrency for processing file system objects
within a pass, fully isolating per-pass shared data structures and global data
structures is not feasible without substantial changes to either the file system
layout or the \CR. As a result, data parallelism does not fully benefit from
increasing CPU count and in fact, as our results show, can degrade substantially
in performance at higher core counts due to increasing synchronization overheads.

To reduce time on synchronization and increase the CPU effectiveness, pipeline
parallelism breaks the limitation that \CR passes must be sequentially executed,
thereby allowing a subsequent C/R pass ($Pass_{i+1}$) to start even before the
completion of an earlier pass ($Pass_{i}$) in a \textbf{pipelined} fashion
(i.e. checking directories in directory checking pass (Pass-2) even before
the inode checking pass (Pass-1) has completed).

\subsubsection{\bf Per-Pass Thread Pools and Work Queues.} First, to facilitate each
pass operating in parallel, we use \textit{per-pass thread pools}. As shown in
Figure~\ref{datafull}, the inode and directory checking passes each maintain a
separate thread pool that is used to hold threads that carry out logic within
the pass. In addition to the per-pass thread pools, each pass maintains a dedicated work
queue filled with file system objects needing to be checked. As each pass
operates, any intermediate work generated is placed in the next pass's work
queue. For example, within the inode checking pass (Pass-1) as directory inodes
are identified, their directory blocks are queued to the directory checking pass's
work queue so they can be checked.

\subsubsection{Overcomming Dependent Checks with Delayed Certification.}
Allowing multiple passes to run in parallel using pipeline parallelism requires
reordering logical checks for correctness. Take an example of the inode checking
pass (Pass-1) and the directory checking pass (Pass-2): in \sysname (and \fsck), the
inode checking pass reads inodes from disk and checks all inodes including directories
and adds directory block information in a shared directory block list
(\emph{dblist}) so the directory checking pass can check the directory entries. While the inode and
directory checking passes can proceed in parallel, the directories can be marked
as consistent only after the inode checking pass verifies the consistency of the inodes representing
the files and subdirectories inside a directory.

\noindent{\bf Providing Ordering Guarantee. }
To address the challenge of ordering guarantee, \sysname delays certain checks
until the prior pipeline pass is complete.  For example, the inode checking pass
within the pipeline is responsible for creating in-memory directory structures
that are used in the directory checking pass to check directories. The directory
checking pass stores a list of subdirectories and checks whether the
subdirectory's parent entry point (represented by double dot $..$) map back to
the directory. However, because the inode checking and directory checking passes
run in parallel, not all the inodes of the directory entries would have been
checked when the parent directories are checked in the directory checking pass.
To handle such scenarios, \sysname delays certification by adding the
uncompleted checks, just like the one described, in a separate work queue and
completing them only after all inodes have been checked (e.g. after the inode
checking pass completes) as shown in Figure~\ref{datafull}.

\subsubsection{Reducing I/O Wait Time in Pipeline Parallelism.}
Effective use of multiple CPUs for an I/O-intensive \CR requires efficient I/O
prefetching and caching even for fast modern storage devices such as NVMe.
Though current \CRs such as \fsck cache and prefetch file system blocks, they are
inflexible and lack thread awareness. For example, \fsck by default prefetches
a few blocks at a time when reading in inodes and fetches only 1 block when fetching
directory blocks. It is possible to change the amount of readahead done however
we observe that statically or naively increasing the
prefetch depth negatively impacts performance because threads access the file
system blocks at different offsets, frequently invalidating previously read
cache entries, consequently increasing the overheads of I/O.

\par To overcome such limitations and accelerate I/O, we implement a per-thread
caching and readahead-based prefetching mechanism that prevents the eviction of cache
entries when multiple threads operate in parallel.
As our results show in Section~\ref{sec:eval}, combining pipeline parallelism
with data parallelism and employing dynamic workload-based threading (discussed
next) improves \sysname's performance across different file system
configurations.

\subsection{Dynamic Thread Scheduling in \sysname}
\label{dynamicsched}
The runtime of C/Rs can vary significantly depending on the configuration of the
file system. For example, C/R on a file system with a larger ratio of smaller
files could result in a substantially longer runtime compared to a file system with
few, but large files due to more metadata needing to be checked. Similarly,
heterogeneity in terms of inode types (files, directories, links) can impact
runtime, and the exact configuration remains unknown until the inodes are iterated
over in the inode checking pass (Pass-1). Additionally, each pass within C/R have
differing degrees of accesses to shared structures. Therefore, a static
assignment of threads across each pass could be ineffective.
Hence, to adapt to file system configurations, \sysname implements
a \CR-aware scheduler, \textbf{\pfscks}, supported by extending the thread pools
to allow for migration of threads between the passes. In addition, \pfscks
maintains an idle thread pool to hold any threads not scheduled to run for any of
the passes.

\noindent{\bf Thread Assignment and Migration of Worker Threads. }
In \sysname, we enable dynamic assignment of threads across each pass by implementing a
scheduler that actively monitors progress and migrates threads across the
passes. The scheduler periodically scans through the work queues of each pass
to identify the work distribution ratio across the pipelined passes and uses
this ratio to assign threads across them.

\begin{figure}
 \hspace{-0.1in}
 \includegraphics[width=8.5cm]{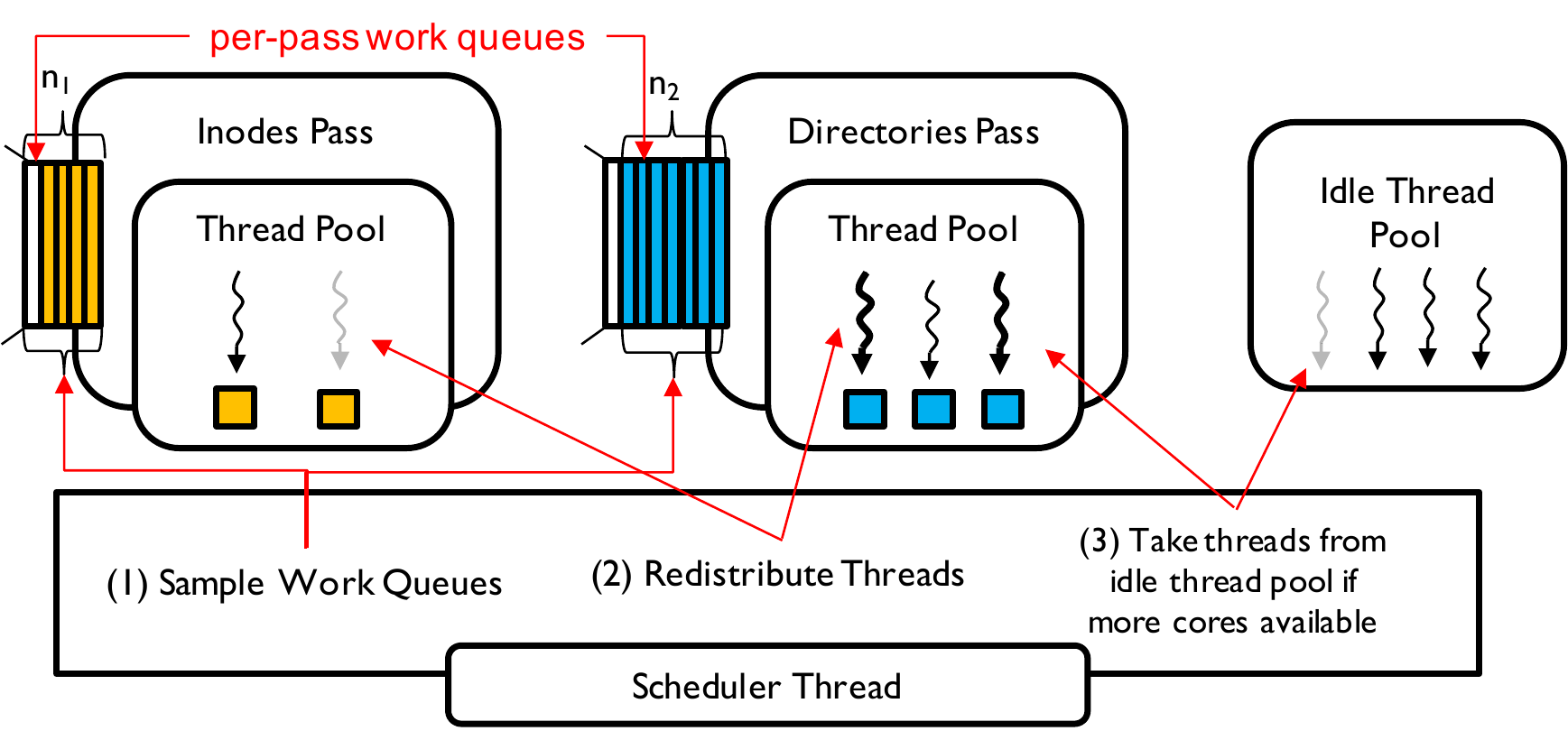}
  \caption{\textbf{Dynamic Thread Scheduling.} A dedicated scheduler threads periodically samples
  work queues among all the passes and redistributes threads based of proportion of outstanding work.
  }
 \label{fig:scheduler}
\end{figure}

Figure~\ref{fig:scheduler} shows an example of \pfscks across the
first two passes. Initially, all the CPU threads are assigned
to the first pass (inode checker) given that \sysname only knows total inodes
from the file system superblock and not the types of inodes. When the inode
checker's worker thread identifies a group of directory inodes, it places the directory
inodes and their corresponding directory blocks to the work queue of directory
checking pass. If no threads are present in the thread pool used for the directory
checking pass, threads from the inode checking pass (first pass) are migrated to
the directory checking pass. To calculate the number of threads to be
reassigned, a dedicated scheduler thread finds the total work to be done
across all passes using the following model.

Let $W_{total}$ be the amount of work needing to be done. Let $q_i$ be the length of the
work queue for pass $i$. Let $n_i$ be the number of discrete elements needing
to be processed for each entry in the work queue. Let $w_i$ be some weight
that normalizes the work to be done for each element in pass $i$. Let $C$ be
the core budget and $t_i$ be the number of threads to assign for pass $i$.
\vspace{-0.1in}
\begin{equation}
  \label{eqn:totalwork}
  W_{total} = \sum_{i=0}^N q_in_iw_i
\end{equation}
\vspace{-0.1in}
\begin{equation}
  \label{eqn:corebudget}
  t_i = C \cdot q_in_iw_i \cdot
\frac{1}{W_{total}}
\end{equation}

As shown in Equation~\eqref{eqn:totalwork}, the total work needing to be done
is a summation of outstanding work across all the passes. The outstanding work
in each pass is a product of the work queue length ($q_i$), the number of objects
encapsulated within each queue entry ($n_i$), and a normalizing weight ($w_i$).
As shown in Equation~\eqref{eqn:corebudget}, with the total amount
of work needing to be done, the scheduler can determine the ideal number of threads to
assign to a pass ($t_i$) based on the total core budget ($C$) and the relative
amount of work calculated for each pass. Note that the normalizing weights are essential for
accounting the differences in the time to process different file types (directories vs. regular files).
In our experimentation, we find it is beneficial to use higher weights for
prioritizing work in the directory checking queue as directories
can take significantly longer to check compared to regular files due to directory
checksum calculations.

\subsection{System Resource-Aware Scheduler}
File system C/Rs could potentially coexist or even share CPUs with other
applications using the same or another file system (or disk). In the pursuit of
exploiting parallelism, \sysname's approaches must avoid or minimize the
performance impact on other applications. To address this goal, we introduce
\textbf{\pfsckr}, which enables resource-awareness for \sysname's scheduler.

\subsubsection{Efficient CPU Sharing}
First, we discuss the case when \pfsckr runs alongside other applications but
using different file systems, where \pfsckr performs C/R on a separate,
unmounted disk. Initially, \pfsckr schedules the main \sysname process using the
SCHED\_IDLE priority to minimize any contention on CPUs with regular processes.
The SCHED\_IDLE priority mostly schedules a process on any idle
CPUs~\cite{schedidle}. As the scheduler periodically runs, \pfsckr first determines a
core budget that represents the maximum number of threads \pfsckr should be
running at any point in time. It does this by identifying the number of CPU
threads currently running, the number of idle cores available, and the number of
cores \pfsckr is currently running on. For idle cores not being utilized by any
application (including \sysname), \pfsckr increases the core budget by the number
of idle cores available. On the contrary, if \pfsckr identifies that the total
number of \sysname threads are more than the idle cores, \pfsckr reduces the
core budget to avoid multiplexing \sysname's threads on the cores it runs on. The
core budget remains unchanged if the available idle cores and \sysname threads
remain the same. After determining the core budget, the scheduler identifies the
work ratio across passes to determine the ideal number of threads that should be
assigned to each pass. The scheduler then redistributes the threads across the
thread pools. In the case of the threads needing to be added due to an increase
in the core budget, threads are taken from the idle thread pool and assigned to
the thread pool to fulfill the ideal thread count. If threads need to be removed
due to a decrease in the core budget, threads are signaled and reassigned to the
idle thread pool. Our results in Section~\ref{sec:eval} show the performance
benefits and implication of \pfsckr when co-running and sharing CPUs with
another application (RocksDB).

\subsubsection{Efficient CPU and File System Sharing}
\label{section:cpuandfssharing}
Given the renewed focus on supporting online-checking, open-source C/Rs such as
\fsck support C/R on a online file system (and disk) that is mounted and actively
used by applications.  \Fsck supports online-checking by utilizing Linux's Logical
Volume Manager's (LVM) snapshot feature which preserves a file system's state by
capturing the changes to the file system ~\cite{hasenstein2001logical}. \Fsck can then perform C/R on
snapshot. The C/R time is dominated by an application's activity in changing the file system objects.
Consequently, this results in a longer C/R time for an actively modified
file system. \\
\indent Despite this, \sysname still shows general improvement over \fsck.
First, \sysname's generic fine-grained parallelism supports and
accelerates online C/R even when applications are sharing the same file system
(and disks). Second, \sysname's resource awareness reduces the impact on co-running
application. As our results show, even when running online C/Rs with
I/O-intensive applications (RocksDB~\cite{RocksDB}), \sysname provides
considerable performance gains compared to running online C/R with vanilla
\fsck.
\subsection{Verifying Correctness and Optimzations}
\noindent{\bf Correctness. }
To ensure correctness with fine-grained \CR parallelism, \sysname employs a
series of steps. First, although the checks are done in parallel, an inode is
not marked complete unless prior passes in the pipeline complete. For example, recall
that a directory within the directory checking pass (Pass-2) cannot be marked as complete
until all the inodes for all of its directory entries are checked. Second, threads
are synchronized when performing complex fixes upon detecting errors. When a worker thread
detects any inconsistencies, all threads across the different passes are notified and
stalled using a barrier. The thread that detected the inconsistency attempts to
fix the errors with (e.g., incorrect inode, blocks claimed by multiple inodes)
or without user input  (e.g., inconsistent bitmap), after which parallel
execution is resumed.
In addition, we plan to explore \CR crash-consistency for \sysname in the future using prior
approaches~\cite{Gatla:Robust}.

\noindent{\bf Optimizations. } As additional optimizations to both \fsck and
\sysname, we restrict the overheads of language localization discussed
earlier to inode and directory block checking, use Intel hardware-accelerated CRC
instead of the default CRC, as well as improve the cache-readahead mechanism.
While the language localization optimization has been reported and up-streamed
to \fsck codes mainline, the other optimizations are under review. We evaluate
the benefits of these optimizations in Section~\ref{sec:eval} (referred to as
\fsck-opt in graphs).

%% file: sections/6_evaluation.tex
\section{Evaluation}
\label{sec:eval}
\vspace{-0.1in}

\begin{table}
\small
\begin{center}
\begin{tabular}{ | m{5em} | m{5.5cm}| }
\hline
Name & Description \\
\hline
\hline
\fsck & original FSCK for EXT file systems \\
\hline
\fsck-opt & optimized \fsck\\
\hline
xfs\_repair & XFS file system checker \\
\hline
\sysname & proposed file system checker \\
\hline
\end{tabular}
 \vspace{-0.05in}
\caption{\bf \CR systems evaluated.}
\label{tab:table-nam1}
\end{center}
\end{table}

\begin{table}
\vspace{-0.4in}
\footnotesize
\begin{center}
\begin{tabular}{ | m{3.6cm} | m{4.1cm}| }
\hline
Name & Description \\
\hline
\hline
datapara & Only data parallelism enabled\\
\hline
datapara+\pipelinespliteq & Pipeline parallelism combined with data
  parallelism, using an equal distribution of threads across the passes\\
\hline
datapara+\pipelinesplitbest & Pipeline parallelism combined with data
  parallelism, using optimal manual thread assignment \\
\hline
sched & All prior parallelization optimizations but using dynamic thread scheduler for thread assignment \\
\hline
rsched & Sched configuration with resource-awareness enabled\\
\hline
\end{tabular}
\vspace{-0.1in}
\caption{\label{tab:table-name} \bf \sysname incremental system design}
\vspace{-0.1in}
\end{center}
\end{table}

Our evaluation of \sysname aims to answer the following
important questions:
\begin{tightitemize}
\item Does data parallelism improve runtime performance by increasing CPU parallelism?
\item Can does pipeline parallelism address the limitations of data parallelism by running multiple passes of \CR simultaneously?
\item How effective is \sysname's dynamic thread placement mechanism for different file system configurations?
\item Can \sysname's resource-aware scheduler effectively minimize the performance impact on other applications when sharing CPUs?
\item How does \sysname perform for online file system checking?
\end{tightitemize}

\subsection{Experimental Setup}
We use a \nvmesysconfig. We run \sysname on various file system configurations
with varying thread counts. As seen in Table~\ref{tab:table-nam1}, we compare
against vanilla \fsck, \fsck-opt (an optimized version of \fsck that
removes localization overheads and utilizes Intel CPU-accelerated CRC calculations),
 and finally, \xfsck. Table~\ref{tab:table-name} shows the incremental \sysname's design
optimizations.

\subsection{Data Parallelism}
In order to evaluate the potential performance improvement with data
parallelism, we run \sysname with just data parallelism (bars shown as
{pFSCK[datapara]}) that parallelizes each pass of a \CR by partitioning
work. We compare two file system configurations: a file-intensive (99\% files)
and directory-intensive (50\% directories) configuration. The x-axis shows the
reduction in runtime with \sysname and the stacked bars shows the runtime
breakdown for each pass.

\begin{figure}[t]
 \includegraphics[width=8cm, keepaspectratio]{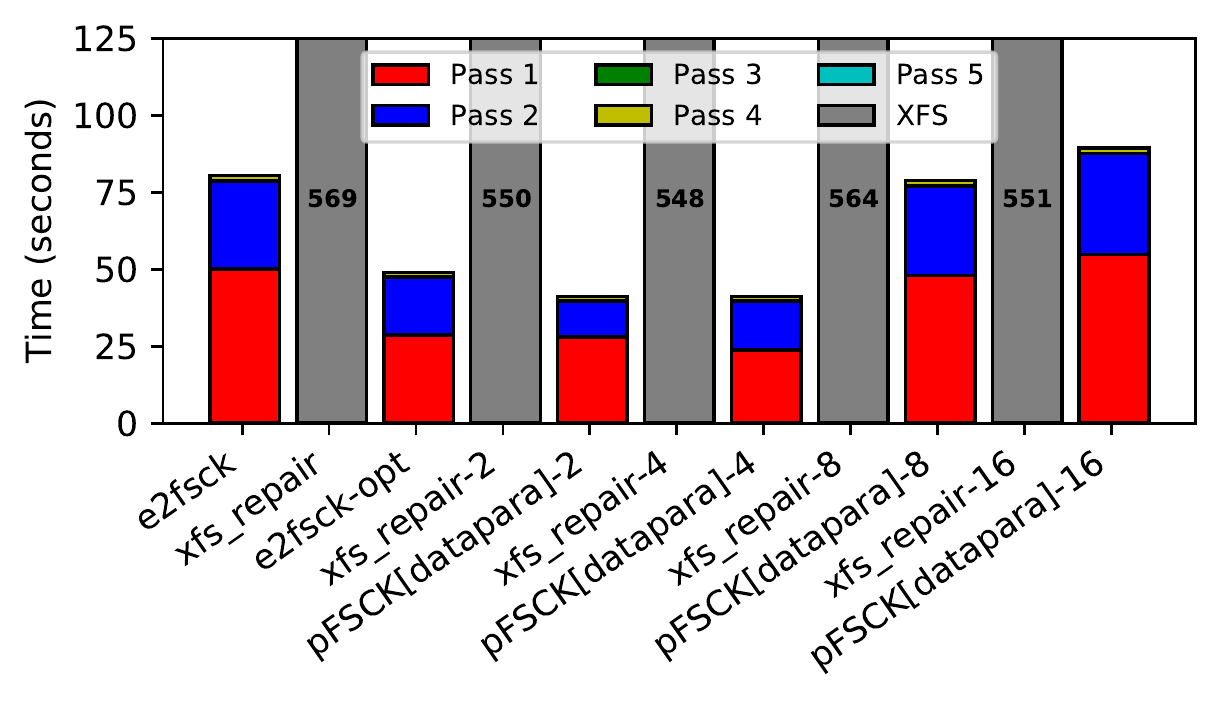}
  \vspace{-0.1in}
  \caption{{\textbf{Data Parallelism impact on a file-intensive configuration.}
  \footnotesize{X-axis shows different \CRs and different thread
  configurations.}}}
\label{fig:datafiles}
\end{figure}

First, with a file-intensive configuration, as shown in
Figure~\ref{fig:datafiles}, the inode checking pass (Pass-1) shows a higher
runtime compared to other passes. Secondly, our optimized \fsck (\fsck-opt)
outperforms the vanilla \fsck by optimizing the CRC mechanism, avoiding language
localization overheads for every inode, and improving the readahead mechanism.
Interestingly, both \fsck and \sysname outperform xfs\_repair for all cases.
Although xfs\_repair is able to check inodes in parallel on the granularity of
allocation groups, it is unable to check directory entries and link counts in
parallel which unfortunately dominates the checking time for large file
systems.

Finally, enabling data parallelism within \sysname reduces the runtime of the
first pass by \pfsckdataparallelfilepassoneperf with 4 threads. Data parallelism
also reduces the runtime of the directory checking pass (Pass-2) by
\pfsckdataparallelfilepasstwoperf resulting in an overall C/R speedup of
\pfsckdataparallelfileperf. Beyond 4-threads, data parallelism does not scale
due to higher serialization and lock contention overheads. Specifically, we find
the functions that update shared structures such as the used/free block bitmap as
the most prominent source of bottlenecks that hinders scaling.

\begin{figure}[t]
 \center
 \includegraphics[width=8cm, keepaspectratio]{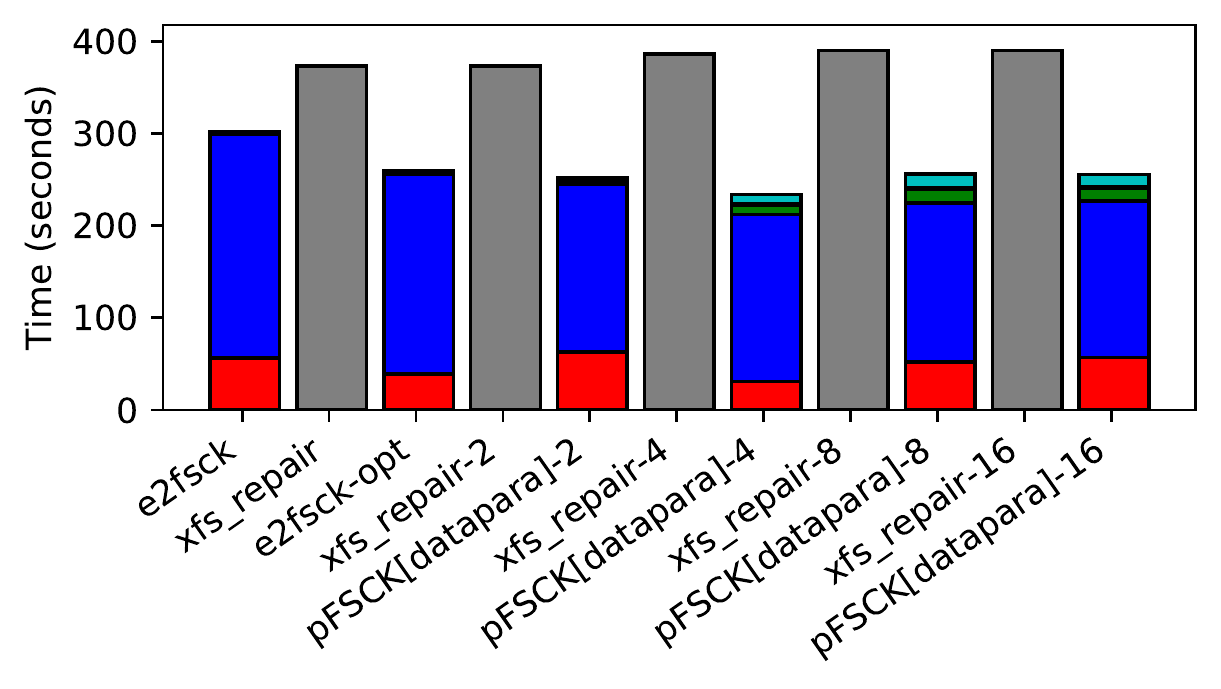}
  \vspace{-0.15in}
  \caption{{\textbf{Data Parallelism impact on a directory-intensive
  configuration.}\footnotesize{The x-axis shows different \CRs and different thread
  configurations.}}}
\label{fig:datadirectories}
\end{figure}

Next, for the directory-intensive file system, as shown in
Figure~\ref{fig:datadirectories}, \sysname parallelizes inode checking (in
Pass-1) and directory checking (in Pass-2). The runtime of Pass-1 and Pass-2
reduces by \pfsckdataparalleldirpassoneperf and \pfsckdataparalleldirpasstwoperf
respectively, resulting in an overall C/R speedup of \pfsckdataparalleldirperf
compared to the vanilla \fsck. Similar to the file-intensive configuration, we
see a performance gain of \totalpfsckoverxfs compared to xfs\_repair, for which
the directory metadata check is not parallelized. Finally, \sysname's data
parallelism does not scale beyond 4-cores.

\begin{figure*}[t]
    \captionsetup[subfigure]{justification=centering}
    \vspace{-0.1in}
    \begin{subfigure}[t]{0.32\linewidth}
     \includegraphics[width=5.5cm, keepaspectratio]{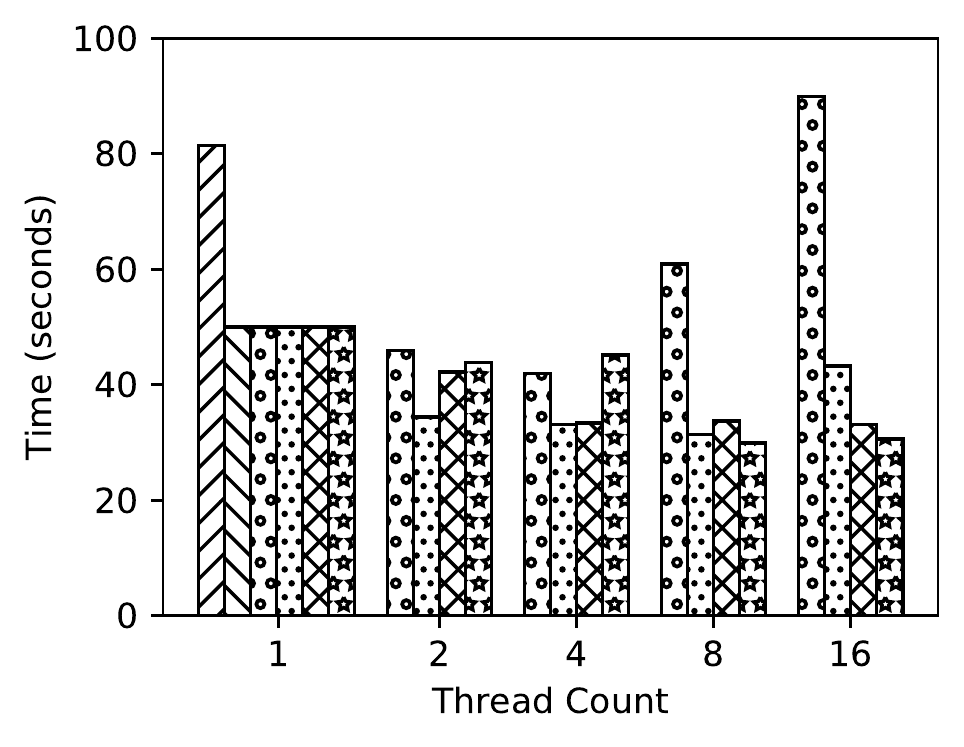}
      \vspace{-0.1in}
      \caption{\bf File-intensive FS on NVMe}
       \label{fig:datapipelinefiles}
    \end{subfigure}
    \begin{subfigure}[t]{0.32\linewidth}
     \includegraphics[width=5.5cm,keepaspectratio]{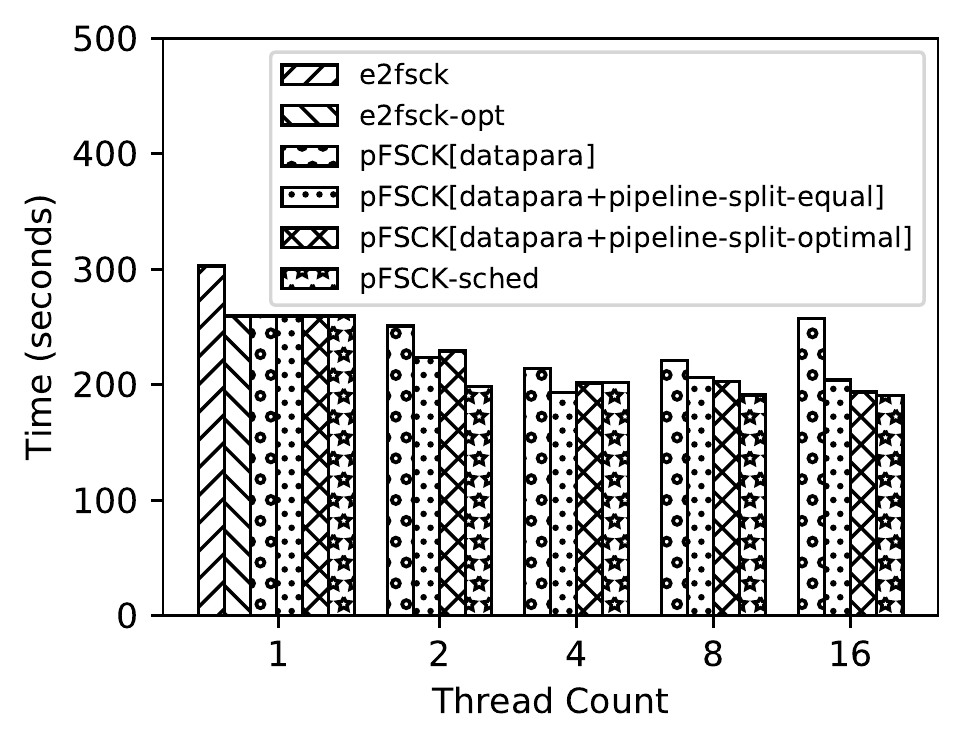}
      \vspace{-0.1in}
      \caption{\bf Directory-intensive FS on NVMe}
       \label{fig:datapipelinedirs}
    \end{subfigure}
    \begin{subfigure}[t]{0.32\linewidth}
     \includegraphics[width=5.5cm,keepaspectratio]{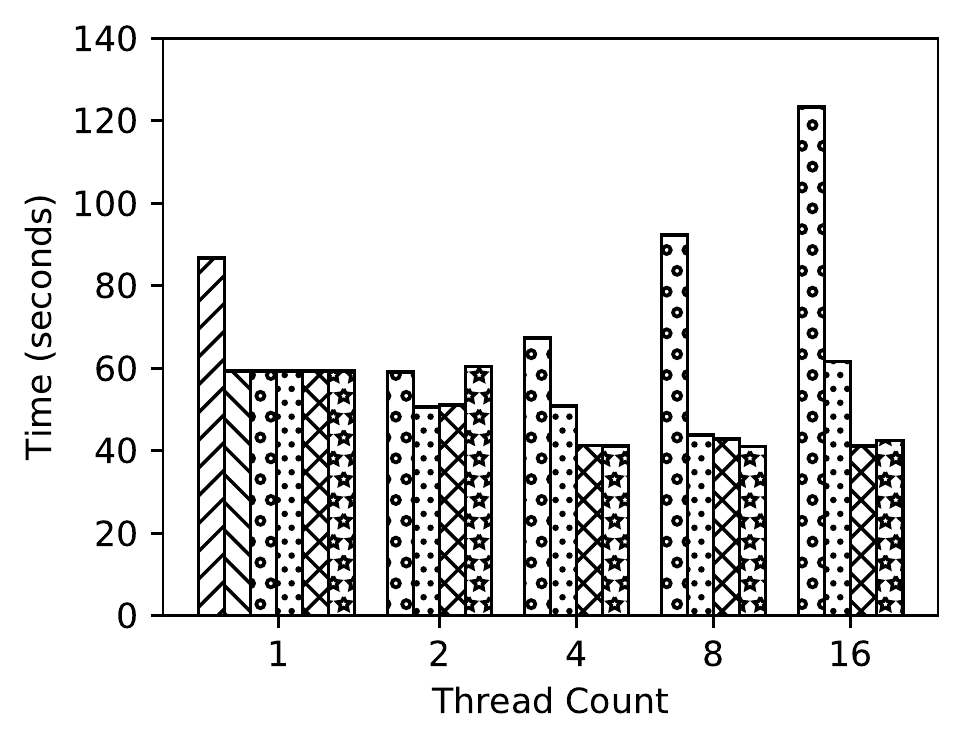}
      \vspace{-0.1in}
      \caption{\bf File-intensive FS on SSD}
       \label{fig:datapipelinefilesssd}
    \end{subfigure}
    \vspace{-0.15in}
    \caption{Comparison of pipeline parallelism and scheduler}
    \vspace{-0.15in}
\end{figure*}

\subsection{Pipeline Parallelism and Dynamic Thread Placement.}
We next evaluate the benefits of combining data and pipeline parallelism and the
need for a dynamic thread placement. In order to evaluate the performance
improvement of pipeline parallelization, we compare \sysname's data parallelism
(\pfsckdatapara), with data and pipeline parallelism
(\pfsckpipeline) against file-intensive and directory-intensive
file system configurations. When using pipeline parallelism, the threads of each
pass adds work to be processed in the next pass. Because of this, logical
boundaries between the passes within \fsck diminishes, hence, we only report the
full runtime of \CR. Because \fsck and \sysname outperforms xfs\_repair in all
cases, we do not show xfs\_repair's results.

Figure~\ref{fig:datapipelinefiles} and Figure~\ref{fig:datapipelinedirs} show
the results for a file-intensive and directory-intensive file system
configuration. The x-axis shows the increase in the number of threads used for
the \CR. In the figures, we compare four cases: (1) \pfsckdatapara, which only
uses data parallelism running one pass at a time, (2) \pfsckpipelinespliteq,
which combines pipeline and data parallelism and statically uses an equal number
of threads for each of the simultaneously executing passes (ex. 2 threads are
assigned to Pass-1 and 2 threads to Pass-2 in a 4-thread configuration), (3)
\pfsckpipelinesplitbest, which represents a best manually selected thread
configuration, and (4) pFSCK-sched,
which dynamically assigns threads to each pass based on the amount of work to be
done and current progress.

First, for the file-intensive configuration, \pfsckpipelinespliteq is not
beneficial for low thread counts (2 and 4 threads) compared to the data
parallelism-only approach, because threads are statically and equally assigned
to Pass-1 and Pass-2 irrespective of the file system configuration. Increasing
the thread count improves performance, because, for a file-intensive
configuration, most work is done in the inode checking pass (Pass-1). As a result,
threads statically assigned for subsequent passes are under-utilized.
Increasing the thread count (along the x-axis) improves \sysname's performance
marginally compared to using only the data parallelism. In contrast,
\pfsckpipelinesplitbest, the manually selected thread configuration case
improves performance by up to \pfsckdatapipelinefileperfvsdata compared to data
parallelism by employing three-fourth of the threads to the inode checker. More
importantly, the manually selected configuration avoids high synchronization
and contention overheads of the data parallelism beyond four threads. Finally,
\sysname's scheduler (\sysname-sched) automatically migrates threads based on
relative amount the outstanding work to be completed across each pass. As a
result, \sysname-sched can automatically provide optimal performance,
improving performance by \pfsckschedfileperfvspipe compared to
\pfsckpipelinesplitbest, resulting in an overall speedup of \pfsckschedfileperf
compared to vanilla \fsck.

\par Next, for the directory-intensive file system, unlike the data parallelism only
approach, pipeline parallelism, which enables simultaneous inode and directory
checking (without compromising correctness), provides some performance
improvement as well. \pfsckpipelinesplitbest reduces runtime by
\pfsckdatapipelinedirperfvsdata with up to 8 threads compared to just using data
parallelism.  When employing \sysname's scheduler (pFSCK-sched), it
automatically mitigates thread contention within the directory block checking by
limiting the number of threads assigned to each pass. Consequently, the
performance improves by \pfsckschedirperfvspipe over \pfsckpipelinespliteq,
resulting in an overall speedup of up to \pfsckscheddirperf over the vanilla
\fsck. Compared to the speedup with a file-intensive file system, since \sysname
employs delayed certification of directories, beyond 8 threads, processing
dependent directories (directory with subdirectory) must join, synchronize, and
merge their work for correctness, limiting scalability.
\par Interestingly, although not shown, we see that \fsck uses the most memory for
directory-intensive file systems. In order to check an 800GB directory-intensive file system,
\fsck uses as much as 3GB of memory as a significant amount of in-memory data structures
are needed to track all the directory information needed to verify the relationships between
the directories and the files and subdirectories that exist within them. Despite this, we find \sysname's memory usage
is comparable, only using as much as 3.5GB resulting in only a \totalpfsckmemoverhead increase in
memory usage.
\par Lastly, in ~\ref{fig:datapipelinefilesssd} shows \sysname performance on
a file-intensive file system backed by SSD.
We see that \sysname is able to show the similar speedups of up to \pfsckschedperfSSD
over vanilla \fsck despite SSDs having lower bandwidth capabilities compared to
NVMe.
\textit{In summary, \sysname's pipeline parallelism reduces the serialization
bottlenecks of data parallelism, and the dynamic thread placement reduces work
imbalance, all leading to significant performance gains.}

\subsection{System Resource-Aware Scheduler}
We next evaluate the effectiveness of \sysname's resource-aware scheduler (\sysname-rsched) in
reducing the impact on other applications compared to \fsck. To illustrate the effectiveness, we
pick a popular persistent key-value store, RocksDB~\cite{RocksDB}, and use it to
run a multithreaded system workload along with each system. We evaluate them in
both an offline setting, where the checker and RocksDB operate on separate file systems,
and an online setting, where the checker and RocksDB operate on the same file system.
For each setting, we consider the following cases:
(1) \textit{e2fsck-no-cpu-sharing}, where \fsck is runs with RocksDB without
sharing the same CPU cores, (2) \textit{e2fsck-cpu-sharing}, where \fsck runs with
RocksDB while sharing the same CPU cores, (3) \textit{\sysname-rsched-no-cpu-sharing}, where
\sysname-rsched runs with RocksDB without sharing the same CPU cores, and
(4) \textit{\sysname-rsched-cpu-sharing}, where \sysname-rsched runs with RocksDb while sharing
the same CPU cores. We run RocksDB with 12 threads. In the case of CPU sharing we force \fsck to share cores
with RockDB by restricting the affinity of all the threads to 12 cores resulting in the
overlapping of one core. In the case of CPU sharing with \sysname-rsched, we run \sysname-rsched with
12 threads and restrict the affinity of all threads to 16 cores, resulting in the overlapping
of 8 cores. For brevity, we show only the results for checking a file-intensive file system.

\subsubsection{Offline \CR with CPU Sharing.}
Figure~\ref{fig:awarefiles} shows the performance of \fsck and \sysname-rsched
when sharing CPU cores with RocksDB. In this experiment, the \CR and RocksDB do not share the file
system, and the \CR runs on an offline unmounted disk. In the y-axis, the
results are normalized relative to no-CPU sharing between the \CR and RocksDB as
the baseline.

\par First, sharing CPUs between \fsck and RocksDB impacts the performance of
both \fsck and RocksDB (shown with e2fsck-cpu-share in the x-axis) compared to
the baseline that does not share CPUs (e2fsck-no-cpu-share). When sharing CPUs,
\fsck's performance degrades by \fsckfsckdeg and RocksDB's performance degrades
by \fsckrocksdeg compared to the baseline. \Fsck is context
switched to run periodically, taking away effective CPU time from RockDB and
introducing overhead from context switching.

\indent Next, for \sysname, the baseline pFSCK-rsched-no-cpu-share configuration shows the
performance of \sysname and RocksDB without CPU sharing, whereas
pFSCK-rsched-cpu-share shows the performance with CPU sharing. For CPU sharing,
we see that although we overlap 8 out of 16 cores, the performance of
\sysname-rsched and RocksDB does not degrade as significantly compared to the
no-sharing case. We see that the performance of \sysname-rsched and RocksDB
degrades only by \pfsckpfsckdeg and \pfsckrocksdeg, respectively; this is
because of \sysname-rsched's ability to downscale the number of threads being
utilized to carry out \CR mitigates the event of CPU time being taken away
from RocksDB and minimizing any potential overhead due to context switching.
Because RocksDB mainly utilizes 12 out the 16 cores for the majority of
execution, the effective performance of \sysname-rsched is equivalent to the
performance of only utilizing 4-6 threads

\subsubsection{Online \CR with CPU Sharing.}
\begin{figure}
\centering
\vspace{-0.2in}
  \begin{subfigure}[t]{0.23\textwidth}
  \hspace{-0.3in}
  \includegraphics[width=4.2cm, keepaspectratio]{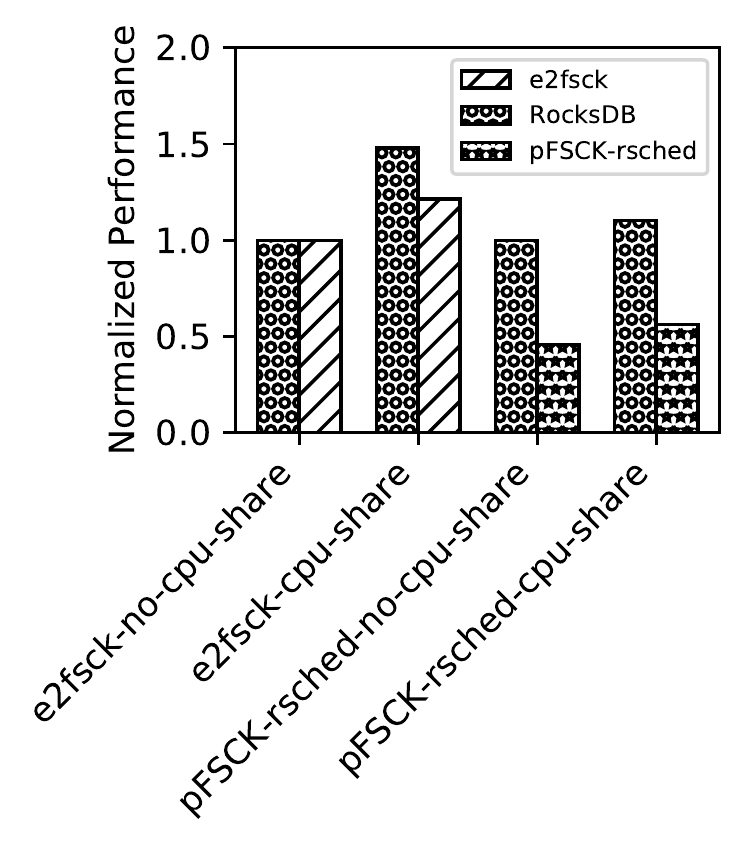}
  \vspace{-0.12in}
  \centering
  \caption{\bf Offline \CR.}
  \label{fig:awarefiles}
\end{subfigure}
  \begin{subfigure}[t]{0.23\textwidth}
  \centering
  \hspace{-0.3in}
  \includegraphics[width=4.2cm, keepaspectratio]{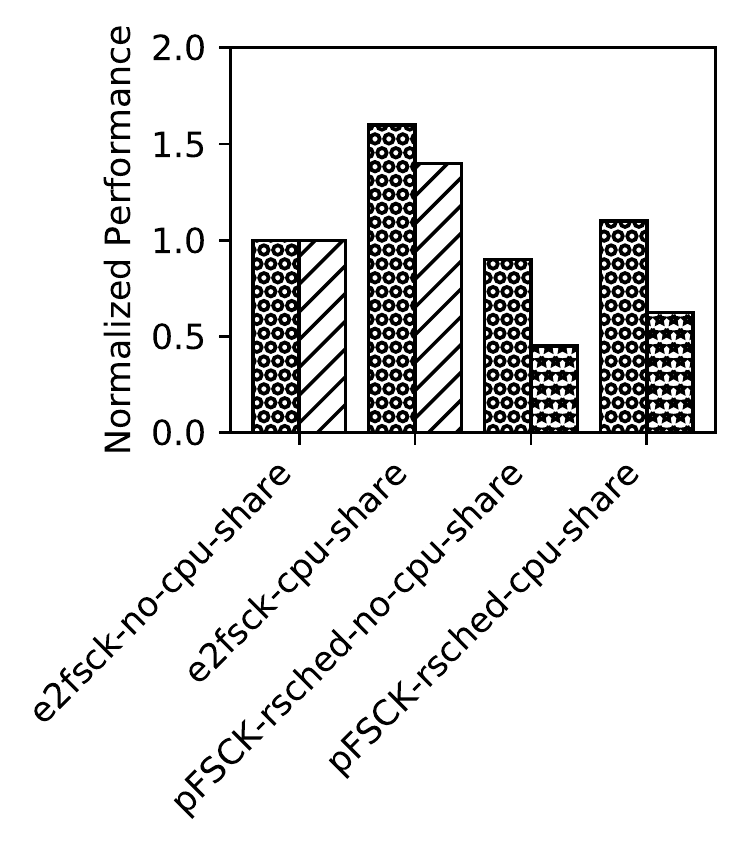}
  \vspace{-0.12in}
  \caption{\bf Online \CR.}
  \label{fig:onlineawarefiles}
\end{subfigure}
  \vspace{-0.1in}
  \caption{{\bf Impact of resource-aware \sysname for offline and online \CR.}
  \small{Results shown for file-intensive configuration.}}
\end{figure}

Figure~\ref{fig:onlineawarefiles} shows the results when \CR and RocksDB share
the CPU as well as the file system. As discussed earlier in
Section~\ref{section:cpuandfssharing}, \sysname utilizes the LVM-based snapshots
to capture the changes to the file system and perform incremental \CR. Similar
to our evaluation with separate file systems, we normalize all the results to
the base of \fsck running with RocksDB without overlapping cores.

First, when overlapping \fsck and RocksDB, performance significantly degrades by
\fsckfsckdegonline and \fsckrocksdegonline respectively.  Interestingly this is
not only due to context switching between \fsck and RocksDB but also due to
overheads in utilizing LVM snapshots. As discussed before, LVM preserves file
system state for a snapshot by capturing all updates to a file system and making
a copy of the original data when modified for the entire duration the LVM
snapshot is active. In the case of \fsck sharing CPUs with RocksDB, since CPU
sharing and context switching naturally decreases the performance of \fsck, this
means the snapshot is active for a longer period of time, compounding the
degradation of performance for both \fsck and RocksDB.

Despite the compounding performance degradation due to LVM snapshot overheads,
the performance degradation when co-running \sysname-rsched with RocksDB is
minimal due to the following reasons: (1) \sysname-rsched's parallelization
reduces \CR runtime compared to \fsck, minimizing the amount of time the
snapshot is active and reducing performance degradation for \CR and RocksDB. (2)
In the case of overlapping cores, similar to an offline setting, \sysname-rsched
mitigates significant performance impact by scaling the number of threads it
uses, reducing performance degradation of both \sysname-rsched and RocksDB by
only \pfsckpfsckdegonline. Compared to an offline setting, this increase in
degradation from \pfsckpfsckdeg for \sysname-sched and \pfsckrocksdeg for
RocksDB to \pfsckpfsckdegonline for both \sysname-sched and RocksDB is due to
compounding overheads of using LVM snapshots.

\par\noindent{\bf Summary. } To summarize, \sysname-rsched's resource awareness
is able to effectively adapt its number of threads to maximize the utilization
of available cores (and performance) in both an offline and online setting while
effectively minimizing the amount of impact on RocksDB.

%% file: sections/8_conclusion.tex
\section{Conclusion}
With a goal of accelerating file system checking and repair tools, in this
paper, we propose \sysname, a parallel \CR tool that exploits CPU parallelism and
the high bandwidth capabilities of modern storage to accelerate file system checking and repair
time without compromising correctness. \sysname explores fine-grained parallelism by
assigning threads with inodes, blocks, or directories and efficiently performing
\CR using data parallelism within each pass and pipeline parallelism across
multiple passes. In addition, \sysname also enables efficient thread management
techniques to adapt to varying file system configurations as well as
minimize performance impact on other applications. Evaluation of \sysname
shows more than \totalpfsckoverfsck gains over \fsck and \totalpfsckoverxfs over \xfsck that provides
coarse-grained parallelism.